\documentclass[12pt]{article}
\newtheorem{Definition}{Definition}

\newtheorem{Theorem}{Theorem}

\newtheorem{Lemma}{Lemma}
\newcommand{\calA}{${\cal A}$}
\newcommand{\calM}{${\mathcal{M}}$}
\newcommand{\calMM}{${\mathcal{M}}_0$}
\newcommand{\calG}{${\mathcal{G}}$}
\newcommand{\calAG}{$\mathcal{A}_{{\mathcal{G}}}$}

\begin{document}

\title{General Relativity on Random Operators}

\author{Michael Heller\thanks{%
Correspondence address: ul. Powsta\'nc\'ow Warszawy 13/94, 33-110 Tarn\'ow,
Poland. E-mail: mheller@wsd.tarnow.pl} \\
Vatican Observatory, V-00120 Vatican City State \\
Copernicus Center for Interdisciplinary Studies, Cracow, Poland
 \and Leszek Pysiak \\
Department of Mathematics and Information Science, \\
Warsaw University of Technology \\
Plac Politechniki 1, 00-661 Warsaw, Poland\\
Copernicus Center for Interdisciplinary Studies, Cracow, Poland
\and Wies{\l }aw Sasin \\
Department of Mathematics and Information Science, \\
Warsaw University of Technology \\
Plac Politechniki 1, 00-661 Warsaw, Poland\\
Copernicus Center for Interdisciplinary Studies, Cracow, Poland}
\date{\today}
\maketitle

\begin{abstract}
We present a mathematical structure which unifies mathematical structures of general relativity and quantum mechanics.  It consists of the noncommutative algebra of compactly supported, complex valued functions  \calA , with convolution as multiplication, on a groupoid $\Gamma $ the base of which is the total space $E$ of the frame bundle over space-time $M$. A differential geometry based on derivations of \calA \ suitably generalizes the standard differential geometry of space-time, and the algebra \calA , when represented in a bundle of Hilbert spaces, defines a von Neumann algebra \calM \ of random operators that generalizes the usual quantum mechanics. The main result of the present paper is that there exists a space \calMM \, dense in \calM, that is isomorphic with the algebra \calA . This isomorphism allows us to transfer all differentially geometric constructions, generalized Einstein's equations including, made with the help of \calA \ (and its derivations) to the space \calMM . In this way, we obtain a generalization of general relativity in terms of random operators on a bundle of Hilbert spaces. However, this generalization cannot be extended to the whole of \calM , and this is the main mathematical obstacle, at least in this approach, to fully unify theory of gravity with physics of quanta.
\end{abstract}

\section{Introduction}
The root of difficulty in combining general relativity and quantum mechanics into a one physical theory lies in radically different mathematical structures these two physical theories employ to express the nature of gravity and the nature of quantum phenomena, respectively. Gravity resides in the geometry of space-time, whereas quantum phenomena express themselves in the action of operators on Hilbert spaces. In this paper, we approach the difficult problem of how to unify these two structures. We do not claim solving the problem. By presenting a strcture that almost reaches this goal, we rather indicate a difficulty blocking the way. It is clear that to obtain the result both structures should be incorporated into a wider mathematical structure that would preserve, or suitably generalize, the present physical meaning of gravity and quanta. We have proposed such a structure in a series of papers. The latter versions of it are presented in \cite{Full,Conceptual}; the first of these papers focuses on mathematical details, whereas the second deals with conceptual issues (in both these papers the reader can find references to earlier works).
In our proposal the unifying structure is provided by the concept of a groupoid related to space-time. It is a groupoid $\Gamma $, the base of which is the total space $E$ of the frame bundle over space-time $M$. Let \calA \ be a noncommutative algebra of compactly supported, complex valued functions on $\Gamma $ with convolution as multiplication. It turns out that, on the one hand, a differential geometry based on derivations of \calA \ suitably generalizes the standard differential geometry of space-time and, on the other hand, the algebra \calA \ when represented in a bundle of Hilbert spaces defines a von Neumann algebra \calM \ that generalizes the usual quantum mechanics. The main result of the present paper is that there exists a space \calMM , consisting of random operators, dense in \calM, which is isomorphic with the algebra \calA . This isomorphism allows us to transfer all differentially geometric constructions, generalized Einstein's equations including, made with the help of \calA \ (and its derivations) to the space \calMM . In this way, we obtain a generalization of general relativity in terms of random operators on a bundle of Hilbert spaces. However, this generalization cannot be extended to the whole of \calM , and this is the main mathematical obstacle, at least in this approach, to fully unify theory of gravity with physics of quanta.

In Sect. 2, we study various groupoids related to space-time. Sects. 3 and 4 summarize some aspects of our model unifying general relativity and quantum mechanics; they are included to make the present paper self-contained.
In Sect. 5, we study the relationship between the algbera \calA \ on the groupoid $\Gamma $ and the algebra  \calMM \ of random operators, and formulate the generalized general relativity in terms of \calMM . In Sect. 6, we show how to recover space-time $M$ from \calMM , and argue that the predual of \calM \ can be regarded as a ``virtual spectrum'' on which the geometry of random operators develops. Finally in Sect. 7, we write, as an example, the Einstein operator for the closed Friedman world model in terms of random operators.

\section{Groupoids Related to Space-Time}
Let $\pi _{M}:E\rightarrow M$ be a frame bundle with the Lorentz group $G$ (or some of its subgroups) as its structural group. Space-time $M$ is recovered as the quotient space $E/G$. A fiber $E_{x}=\pi _{M}^{-1}(x)$ over $%
x\in M$ is the set of all local reference frames attached to the point $x$. The group $G$ acts on $E$ to the right, $E\times G\rightarrow E$, along the fibers. We construct a groupoid over $E$, called {\it transformation groupoid}, in the following way.

\[
\Gamma =E\times G=\{\gamma =(p,g):p\in E,g\in G\},
\]
with $\Gamma_0 = E$ as its base. If  $\gamma _{1}=(p_{1},g_{1})$, $\gamma
_{2}=(p_{2},g_{2}) \in \Gamma$, the multiplication is defined
$$\gamma _{1}\circ \gamma _{2}=(p_{1},g_{1}g_{2})$$
provided that $p_{2}=p_{1}g_{1}$.
As the source and range mappings we have
\[ d(p,g)=p, \;\;r(p,g)=pg, \]
respectively. The inverse of $\gamma =(p,g)$ is $\gamma ^{-1}=(pg,g^{-1})$.  The identity section (the set of units) is
defined to be $\Gamma ^{(0)}=\{(p,e):p\in E\}$ where $e$ is the unit of $G$. (For details see \cite{Landsman,Paterson} or our earlier works \cite{Finite,Pysiak2004}.) 88888

Let ${\cal A} =C_c^{\infty}(\Gamma ,{\bf C})$ be the algebra of smooth, compactly supported, complex valued functions on
$\Gamma$. \calA \ is a noncommutative algebra if the multiplication is defined as the convolution
\[
(f_1*f_2)(\gamma )=\int_{\Gamma_{d(\gamma )}}f_1(\gamma_1)f_2(\gamma_1^{-1}\gamma )d\gamma_1,
\]
for $f_1,f_2\in {\cal A}$, where $d(\gamma )=d(p,g)=p$.
\par
The idea is to develop differential geometry in terms of the algebra \calA \ just as the usual geometry on a manifold $M$ can be developed in terms of the algebra $C^{\infty}(M)$. It turns out that the lifting of the algebra $C^{\infty
}(M)$ to the total space $E$ of the frame bundle over $M$, i.e., the set $Z = \pi_M^*(C^{\infty }(M))$ (which is, of course isomorphic with $C^{\infty }(M)$), can be regarded as an ``outer center'' of the algebra $\mathcal{A}$ (the ``real center'' of the algebra \calA \ is empty). In fact, the algebra $\mathcal{A}$ is a module over $Z = \pi_M^*(C^{\infty }(M))$. Indeed, although the functions belonging to $Z$ are not compactly supported, their action on the algebra $\mathcal{A}$, $\alpha : Z \times \mathcal{A} \rightarrow \mathcal{A}$, can be defined in the following way
\[
\alpha(f,a)(p,g) = f(p)a(p,g),
\]
$f \in Z, a \in \mathcal{A}$.  Owing to this fact we can develop a noncommutative geometry based on $\mathcal{A}$ that will be a generalization of the standard geometry on a manifold.

Let now $\Gamma_W = \bigcup_{x \in M} E_x \times E_x := E \times_{M} E$ where $E_x$ is a fiber in $E$ over $x\in M$. This, together with obviously defined groupoid operations (composition law reads: $(p_1,p_2) \circ (p_2,p_3) = (p_1,p_3)$), forms what is called the {\it Whitney groupoid}. If $\gamma = (p,q) \in \Gamma_W$ then there exists $g \in G$ such that $q =pg$, that is to say $\gamma = (p, pg)$, which establishes the isomorphism between transformation and Whitney groupoids. Let us notice, however, that if the frame bundle $\pi_M: E \rightarrow M$ is not locally trivial, the isomorphism does not occur. This happens, for instance, when a singular boundary $\partial M$ is attached to $M$ \cite{Malicious,Anatomy}.

Let us notice that $\Gamma_W$ can also be written in the form
\[
\Gamma_W =\{(x,p_1,p_2):p_1,p_2\in
E\;{\rm a}{\rm n}{\rm d}\;\pi_M(p_1)=(\pi_M(p_2))=x\}.
\]
Let us define the algebra ${\cal A}_W=C^{\infty}(\Gamma_W,{\bf C} )$ with
convolution as multiplication
\[
(\tilde {f}_1*\tilde {f}_2)(x,p_1,p_2)=\int_{E_x}\tilde {
f}_1(x,p_1,p_3)\tilde {f}_2(x,p_3,p_2)dp_3.
\]

The mapping $J:{\cal A}_W \rightarrow {\cal A}$ given by
\[J(f)(\gamma )=f(\pi_M(p),p,pg),\]
for $f\in {\cal A}_W,\,\gamma =(p,g)$, establishes the isomorphism of the algebras $\cal{A}_W$ and \calA  \  \cite{Full}.

Let, as before, $E$ be a $G$-principal bundle over $M$. We  define the Cartesian product ${\cal G} = E \times E$ and introduce in it the groupoid structure in the following way. The composition  
$$[p_1,p_2] \circ [p_3,p_4] = [p_1, p_4]$$
is defined if $p_2 = p_3$. The source and target mappings are
\[ d(q,p) = p, \; \; r(p,q) = q,
\]
respectively. We also have the identity 
$$\epsilon (p) = (p,p),$$
and the inverse
$$(p, q)^{-1} = (q,p).$$
The following sets are naturally defined
\[
{\cal G}_{p} = \{(q,p): q \in E\},
\]\[
{\cal G}^{p}= \{(p,q)]: q\in E\}.
\]
This groupoid is called the {\it (full) pair groupoid} or the {\it coarse groupoid}. It is a transitive groupoid. In our case, it is not that coarse since it inherits quite a rich structure after its base space $E$, the total space of the fiber bundle over space-time $M$.
The transformation groupoid is a subgroupoid of the full pair groupoid.

Let \calAG \ be the algebra of smooth compactly supported, complex valued functions on \calG \ with convolution as multiplication
\[
(a_1 * a_2)(p_1,p_2) = \int_E a_1(p_1,p) a_2 (p,p_2) dp
\]
where $dp$ is the  manifold measure on $E$. Let us notice that the above convolution can also be written as
\[
(a_1 * a_2)(p_1,p_2) = \int_M d\mu(x) \int_G a_1(p_1, p(x)g) a_2 (p(x)g, p_2) dg.
\]

\section{Generalized Differential Geometry}
In \cite{Full} we have constructed a differential geometry based on the algebra \calA \ and its derivatives; here we only summarize the main steps of this construction (see also \cite{Conceptual}).

The set of all derivations of the algebra $\mathcal{A}$ will be denoted by $%
\mathrm{Der}(\mathcal{A} )$. It has the algebraic structure of a Z-module.
The pair $(\mathcal{A},V)$, where $\mathcal{A}$ is (not necessarily commutative) algebra and $V\subset
\mathrm{Der}(\mathcal{A})$ is a (sub)module of its derivations, will be called {\it differential algebra}. We first discuss derivations of \calA .

Let $\bar{X}\in \mathcal{X}(E)$$\bar{X}$ be a vector field on $E$. We assume that $\bar{X}$ is a \emph{right invariant\/},  i.e. $(\mathcal{R}_{g})_{\ast p}\bar{X}(p)=\bar{X}(pg)$ 
for every $g\in G$. The \emph{lifting\/} of $\bar{X}$ to $\Gamma $ is
$$ \bar{\bar{X}}(p,g)=(\iota _{g})_{\ast p}\bar{X}$$
where the inclusion $\iota _{g}:E\hookrightarrow E\times G$ is given by $%
\iota _{g}(p)=(p,g)$. The lifting of a right invariant
vector field $\bar{X}\in \mathcal{X}(E)$ to $\Gamma $ is a derivation of the
algebra $\mathcal{A}$.

If the right invariant vector field satisfies the condition $(\pi_M)_*\bar { X}=0$ it is called a \emph{%
vertical\/} vector field. Such vector fields, when lifted to $\Gamma$, are
derivations of the algebra $\mathcal{A}$, and are called \emph{vertical
derivations}.

With the help of the connection in the frame bundle $\pi
_{M}:E\rightarrow M$ we lift a vector field $X$ on $M$ to $E$, i.e., $\bar{X%
}(p)=\sigma (X(\pi _{M}(p)),\,\pi _{M}(p)=x\in M$ where $\sigma $ is a lifting
homomorphism. This vector field is
right invariant on $E$. We lift it further to $\Gamma $,
\[
\bar{\bar{X}}(p,g)=(\iota _{g})_{\ast p}\bar{X}(p)\in \mathcal{X}(\Gamma ).
\]
It can be shown that this lifting preserves the right invariance property, and is a derivation of the
algebra $\mathcal{A}$. We call it a \emph{horizontal derivation\/} of $%
\mathcal{A}$.

The \emph{inner derivations} of the algebra $\mathcal{A}$, $ \mathrm{Inn}(\mathcal{A})=\{ad(a):a\in \mathcal{A}\}$,
are typical for noncommutative algebras.
The mapping $\Phi (a)=ad(a)$, for every $a\in \mathcal{A}$, establishes the isomorphism between
the algebra $\mathcal{A}$ and the space $\mathrm{Inn}(\mathcal{A})$ as $Z$%
-moduli.

Our generalized differential geometry is based on the differential algebra ($\mathcal{A},V)$
with $V=V_{1}\oplus V_{2}$ where $V_{1}$ and
$V_{2}$ are horizontal and vertical derivations, respectively. The $Z$-module $V_3 = \mathrm{Inn}(\mathcal{A})$ is responsible for ``quantum geometry'' in our model (see \cite{Full} or \cite{Conceptual}).

We now assume that $G$ is a noncompact
and semisimple group (which includes the group $SO_{0}(3,1)$), and as the metric $\mathcal{G}:V\times V \rightarrow Z$ we choose
\[
\mathcal{G}(u,v)=\bar {
g}(u_1,v_1)+\bar {k} (u_2,v_2)
\]
where $u_1,v_{_1}\in V_1,u_2,v_2\in V_2$. The metric $\bar{g}$ is evidently the lifting of the metric $g$ on space-time $M$, i.e., $
\bar {g}(u_1,v_1)= \pi_M^{*}(g(X,Y))$
where $X,Y \in \mathcal{X} (M).$ We assume that the metric $ \bar {k}$ is of the Killing type. 

Now, we define preconnection by the Koszul formula
\[
(\nabla _{u}^{\ast }v)w=\frac{1}{2}[u(\mathcal{G}(v,w))+v(\mathcal{G}%
(u,w))-w(\mathcal{G}(u,v))
\]%
\[
+\mathcal{G}(w,[u,v])+\mathcal{G}(v,[w,u])-\mathcal{G}(u,[v,w]).
\]%
In \cite{Full} we have demonstrated that if $V$ is a $Z$-module of derivations of an algebra $(\mathcal{A},\ast )$ such that $V(Z)=\{0\}$ then, for
every symmetric nondegenerate tensor $g:V\times V\rightarrow Z$, there
exists exactly one connection $g$-consistent with the preconnection $\nabla
^{\ast }$, and this connection is given by 
\[
\nabla _{u}v=\frac{1}{2}[u,v].
\]

For $V_{2}$, we assume the metric is of the Killing type $g(u,v)=\mathrm{Tr}(u\circ v)$.
The Killing form, related to a Lie group, is
\[
\mathcal{B}(V,W)=\mathrm{Tr}(a d(V)\circ ad(W))
\]
where $V,W$ are elements of the Lie algebra $\underline g$ of the group $G$.
In our case, this Killing form $\mathcal{B}$ is nondegenerate (since the group $G$ is semisimple). Let us notice that the tangent space to
any fiber $E_x,x\in M$ at a fixed point $p \in E$, is isomorphic to $\underline g$. This allows us (see \cite{Full}) to write the metric $\bar {k} :V_2\times V_2\rightarrow Z$ in the form
\[
\bar {k}(\bar{\bar {
X}},\bar{\bar {Y}})= \mathcal{B}(\bar X(p), \bar Y(p)).
\]

The trace for the algebra $\mathcal{A}_1$ (which is isomorphic to the
algebra $\mathcal{A}$) is 
\[
\mathrm{Tr}(a)(p)=\int_G a(pg , pg)dg \in Z.
\]

We define the \emph{curvature}, for all  $V_i,\,i=1,2$, in the usual way
\[
\stackrel{i}{R}(u,v)w =\stackrel{i}{\nabla}_ u\stackrel{i}{\nabla}_ v w -%
\stackrel{i}{\nabla}_ v\stackrel{i}{\nabla}_ u w -\stackrel{i}{\nabla}_{
[u,v]}w.
\]

For $i=2$ we have
\[
\stackrel{2}{R}(u,v)w =-\frac 14[[u,v],w].
\]

For $i=1,2$ and every endomorphism $T:V_i\rightarrow V_i$, there exists the
usual trace $\mathrm{Tr}(T)\in Z$, and we can define $\stackrel{i}{R}%
_{uw}:V_i\rightarrow V_ i$ by
\[
\stackrel{i}{R}_{uw}( v)=\stackrel{i}{R}(u,v) w.
\]
Therefore, the \emph{Ricci curvature} is
\[
\stackrel{i}{\mathbf{r} \mathbf{i}\mathbf{c}}(u,w) =\mathrm{Tr}(\stackrel{i}{R}%
_{uw} ),
\]
and the \textit{adjoint Ricci operator} $\stackrel{i}{\mathcal{R}}: V_i
\rightarrow V_i$ 
\[
\stackrel{i}{\mathbf{r} \mathbf{i}\mathbf{c}}(u,w) =\stackrel{i}{\mathcal{G}} (%
\stackrel{i}{\mathcal{R}} (u),w)
\]
where $\stackrel{1}{\mathcal{G}} =\bar {g}$ and $\stackrel{2}{\mathcal{G}}=%
\bar {k}$. If the metric $\stackrel{i}{\mathcal{G}}$ is nondegenerate, there
exists the unique $\stackrel{i}{\mathcal{R}}$ satisfying this equation
for every $w\in V_i$.

The \emph{curvature scalar\/} is
\[
\stackrel{i}{r}=T r(\stackrel{i}{\mathcal{R}} ).
\]

For $V_2$ the usual trace exists, and we readily compute
\[
\stackrel{2}{\mathbf{r} \mathbf{i}\mathbf{c}}(u,w)=\frac 14\bar {k}(u,w)
\]
for every $u,w\in V_ 2$.

We have now all necessary magnitudes to define the \emph{generalized Einstein operator\/}
\[
\mathbf{G} := \mathcal{R} - \frac{1}{2}r \mathrm{id}_V
\]
where $r = \mathrm{Tr}\mathcal{R}$. This allows us to naturally define {\it Einstein's generalized equation\/} $\mathbf{G} = 0$, and to have generalized general relativity. In \cite{Conceptual} we postulated that Einstein' generalized equation has the form of the eigenvalue equation for the Einstein operator
\[
\mathbf{G} - \tau \mathrm{id}_V = 0
\label{EigenEinstein}
\]
where $\tau \in Z$.  This has proved to be an interesting modification. It does not only better fit into the quantum character of the model proposed in \cite{Conceptual}, but also leads to the remarkable result: it turns out that when equation (\ref{EigenEinstein}) is computed for the Friedman cosmological model, the eigenvales $\tau $ produce the correct components of the energy-momentum tensor (in spite of the fact that no enegy-momentum tensor has been assumed in the beginning) together with a suitable equation of state. If one restricts to the submodule $V_1$, one essentialy recovers the standard Einstein equation on space-time. (For details see \cite[Sect. 5]{Conceptual}.) In the following, we shall also discuss equation (\ref{EigenEinstein}). 

 When we restrict to the $V_1$-component of our geometry, we recover the usual space-time geometry (lifted to $\Gamma $).

\section{Algebra of Random Operators}
To construct mathematical structures responsible for quantum physics we define the regular representation
\[
\pi_p:\mathcal{A}\rightarrow \mathcal{B}(\mathcal{H}^p) ,
\]
of the algebra \calA \ in the Hilbert space $\mathcal{H}^p=L^2(\Gamma^ p)$ for every $p\in E$ ($\Gamma^ p$ being the set of all elements of $\Gamma$ ending at $p$), where $\mathcal{B}(\mathcal{H}^ p)$ is the algebra of bounded operators on the Hilbert space $\mathcal{H}^p$. This representation $\pi_p$ is given by
\[
(\pi_p(a)\psi )(\gamma )=\int_{\Gamma_{d(\gamma )}}a(\gamma_1)\psi (
\gamma_1^{-1}\circ\gamma )d\gamma_1
\]
where $a\in \mathcal{A} ,\psi\in \mathcal{H}^p, \gamma ,\gamma_1\in\Gamma$.
Here the Haar system on $\Gamma $ is formed by the measure on the group $G$, transferred to each fiber of $ \Gamma $.

We shall show that every $a\in \mathcal{A}$ generates a random operator $
r_a=(\pi_p(a))_{p\in E}$, acting on a bundle of Hilbert
spaces $ \{\mathcal{H}^p\}_{p \in E}$ where $\mathcal{H}^
p=L^2(\Gamma^p)$. Let us first recall the definition of random operators \cite[p. 50-53]{Connes}

An operator $r_{a}$ is a \emph{random operator\/} if it satisfies the
following conditions
(1) If $\xi_p,\eta_p \in \mathcal{H}^p$ then the function $ E\rightarrow
\mathbf{C}$, given by $E\ni p\mapsto (r_a \xi_p,\eta_p)$,
$a\in \mathcal{A}$, is measurable with respect to
the usual manifold measure on $E$. (2) The operator $r_ a$ is \emph{bounded}, i.e., $ ||r_a|| < \infty $ where $||r_a||=\,\mathrm{e}\mathrm{s} \mathrm{s}\,\mathrm{s}\mathrm{u} \mathrm{p} ||\pi_p(a)||$. Here
``ess sup'' denotes supremum modulo zero measure sets.

It is clear that in our case both these conditions are satisfied. For every $p\in E$, $\pi_p(a)$ is a bounded operator on $\mathcal{H}^p$. Let us denote by $\mathcal{M}_{0}$ the algebra of equivalence classes (modulo equality almost everywhere) of these random operators $r_{a},a\in \mathcal{A}$. (Random properties of this structure have been discussed in \cite{Random}.)

Let us consider the direct integral of Hilbert spaces $\mathcal{H}^p$, 
\[
{\cal H}= \int_{\oplus ,p} \mathcal{H}^p d\mu (p)
\]
that is also a Hilbert space. The operator $\pi(p)(a) = (\pi_p(a))_{p \in E}$ acts on $\mathcal{H} $ in the following way
\[
(\pi(a) \psi )(p) = \pi_p(a)(\psi(p)).
\]
The set $\mathcal{M}=\mathcal{M}_{0}^{\prime \prime } $, where $\mathcal{M}_{0}^{\prime \prime }$ denotes the double commutant of $ \mathcal{M}_{0}$ in $\mathcal{H}$, is a von Neumann algebra. It is called the \emph{von Neumann algebra of the groupoid} $\Gamma $ \cite[p. 52]{Connes}.

\section{Bimultiplicative Algebras}
The following Lemma plays an important role in our further analysis
\begin{Lemma}
The mapping $\pi : \mathcal{A} \rightarrow \mathcal{M}_0$, given by $\pi (a) = (\pi_p(a))_{p \in E}$, is an isomorphism of algebras.
\label{Lemma1}
\end{Lemma}
\noindent
{\it Proof} is given in \cite[Sect. 7]{Conceptual}.

The isomorphism of the above Lemma  can be written, more explicitly, as
\[
\pi : (\mathcal{A},*,+) \rightarrow ({\mathcal{M}}_0,\circ , +)
\]
with $\pi_p(a) \circ \pi_p(b) = (r_a \circ r_b)(p)$.

In the algebra \calMM \ we define the commutative multiplication, $\cdot : {\mathcal{M}}_0 \times {\mathcal{M}}_0 \rightarrow {\mathcal{M}}_0$, by $r_a \cdot r_b = r_{ab}$. Evidently, the mapping
\[
\tilde{\pi }: (\mathcal{A}, \cdot , +, \mathbf{C}) \rightarrow ({\mathcal{M}}_0, \cdot , +, \mathbf{C})
\]
given by $\tilde{\pi }(a) = r_a$ is an isomorphism of commutative algebras. In this case, the algebra \calA \ can be simply denoted by $C_{c}^{\infty }(\Gamma )$.

In this way, we have demonstrated that both \calA \ and \calMM \ are {\it bimultiplicative} algebras, i.e., the algebras with double multiplication. Let us notice that the algebra $(\calA, \cdot , +)$ is a $Z$-module, and the algebra $(\calA, *, +)$ is a $Z$-bimodule. The mapping $\pi $ is also the isomorphism of bimultiplicative algebras 
\[
\pi : (\mathcal{A}, \cdot , *, +, \mathbf{C}) \rightarrow ({\mathcal{M}}_0,\cdot , \circ , +, \mathbf{C})).
\]

\begin{Definition}
A mapping $\mathbf{X}: {\mathcal{M}}_0 \rightarrow {\mathcal{M}}_0$ is said to be a {\em semiderivation} if $\mathbf{X}$ is $\mathbf{C}$-linear and satisfies the Leibniz rule with respect to the commutative multiplication ``$\cdot $''. 
\end{Definition}
The set of all semiderivations of the algebra \calMM \ will be denoted by $\rm{SDer}({\mathcal{M}}_0)$.

\begin{Lemma}
For every semiderivation $\mathbf{X}: {\mathcal{M}}_0 \rightarrow {\mathcal{M}}_0$ there exists a derivation $X \in {\rm{Der}}(C_c^{\infty }(\Gamma ))$ such that $\mathbf{X}(r) = \pi(X(\pi^{-1}(r)))$, and the derivation $X$ is also the derivation of the algebra $C^{\infty }(\Gamma )$.
\label{Lemma2}
\end{Lemma}

\noindent
{\it Proof.} Let us define $X: \mathcal{A} \rightarrow \mathcal{A}$ by $X(a) = \pi^{-1}(\mathbf{X}(\pi(a))$. For $\varphi \in C^{\infty }(\Gamma )$
and any element $\gamma $ of $\Gamma $, we extend $X$ to the algebra $C^{\infty }(\Gamma )$ in the following way
\[
(\tilde{X} \varphi)(\gamma ) = (X a)(\gamma )
\]
where $\varphi |U = a|U$ for an open neighborhood $U$ of $\Gamma $.

We thus have the $C^{\infty }(\Gamma )$-module $\rm{SDer}({\mathcal{M}}_0)$ of semiderivations and the $C^{\infty }(\Gamma )$-module $\rm{Der}(C^{\infty }(\Gamma ))$ of derivations. And the map
\[
\psi : \rm{Der}(C^{\infty }(\Gamma )) \rightarrow \rm{SDer}({\mathcal{M}}_0)
\]
given by $\psi (X) = \mathbf{X}$ where
\[
\mathbf{X}(\pi (a)) = \pi (X(\pi^{-1}(a)),
\]
for every $a \in \mathcal{A}$, is an isomorphism of $C^{\infty }(\Gamma )$-modules.

\begin{Lemma}
A semiderivation $\mathbf{X} \in \rm{SDer}({\mathcal{M}_0})$ is a derivation with respect to the noncommutative multiplication ``*'', if $\mathbf{X}$ is right invariant.
\label{Lemma3}
\end{Lemma}
\noindent
{\it Proof} is given in \cite[Lemma 1]{Full}.

In the following, the set of right invariant derivations of \calM \ will be denoted by $\mathrm{RSDer}({\mathcal{M}}_0)$.

\begin{Definition}
We say that two differential algebras $(\mathcal{A}, V)$ and $(\mathcal{B}, W)$, where $V \subset \rm{Der}(\mathcal{A})$ and $W \subset \mathrm{Der}(\mathcal{B})$, are isomorphic, if \calA \ and $\mathcal{B}$ are isomorphic as (non)commutative algebras, and $V$ and $W$ are isomorphic as the corresponding moduli with respect to the commutative multiplication.
\end{Definition}

Now, we can summarize the above Lemmas in the form of the following theorem.
\begin{Theorem}
There exist the isomorphisms of differential algebras
\[
\tilde{\Pi }: (C^{\infty }(\Gamma ), \mathrm{Der}(C^{\infty }(\Gamma ))) \rightarrow (({\mathcal{M}}_0, \cdot ), \mathrm{SDer}({\mathcal{M}}_0))
\]
and
\[
\Pi : (\mathcal{A}, V) \rightarrow (({\mathcal{M}}_0, * ), \mathrm{RSDer}({\mathcal{M}}_0))
\]
where $V$ denotes the $Z$-module of horizontal and vertical derivations of the algebra \calA .
\end{Theorem}

This is an important theorem. Owing to the isomorphism $\tilde{\Pi }$ the commutative geometry of the groupoid $\Gamma $ transfers to the commutative geometry of the space \calMM , i.e., to the algebra of random operators (with the commutative multiplication). Owing to the isomorphism $\Pi $ the noncommutative geometry of the groupoid $\Gamma $, as determined by the algebra \calA \ and its derivations, transfers to the noncommutative geometry if the space \calMM , i.e., to the algebra of random operators (with convolution as multiplication).
For instance, the Einstein operator $\mathbf{G}: V \rightarrow V$ as defined on $\Gamma $ is transferred into the Einstein operator
\[
\hat{\mathbf{G}}: \mathrm{RSDer}({\mathcal{M}}_0) \rightarrow \mathrm{RSDer}({\mathcal{M}}_0),
\]
or more explicitly, to a derivation $u: \mathcal{A} \rightarrow \mathcal{A}, \; u \in V$ there corresponds the derivation $\hat{u}: {\mathcal{M}}_0 \rightarrow {\mathcal{M}}_0, \; \hat{u} \in \mathrm{RSDer}({\mathcal{M}}_0)$ by
\begin{equation}
\hat{u}(r_a) = r_{u(a)},
\label{derM}
\end{equation}
and the Einstein operator
\[
\hat{\mathbf{G}}: \mathrm{RSDer}({\mathcal{M}}_0) \rightarrow \mathrm{RSDer}({\mathcal{M}}_0),
\]
is given by
\begin{equation}
\hat{\mathbf{G}}(\hat{u})(r_a) = r_{\mathbf{G}(u)a}.
\label{EinstOperator}
\end{equation}
In this way, we can transfer the generalized general relativity as defined on $\Gamma $ to the algebra of random operators.

It is important to realize that we can have a generalized general relativity on the space \calMM , but not on the von Neumann algebra \calM \ that is a typical structure for quantum physics. From this one can see that the stumbling block for having quantum gravity on \calM \ lies in the ``limit elements '' of \calM .

\section{Recovering Space-Time}
In this section, we answer the question: how from the algebras \calA \ and $\mathcal{M}_0$ \ space-time can be recovered? First, we prove that there is the bijection between the spectra of the algebras $(\mathcal{A},\cdot ) $ and $(C^{\infty }(\Gamma ), \cdot)$. The proof of this rather obvious fact is by no means trivial. To produce it we need the following concept.

Let us consider a family $C$ of real valued functions on a set $V$ and endow it with the weakest topology $\tau_C$ in which the functions of $C$ are continuous. A function $f: A \rightarrow \mathbf{R}$, where $A\subset V$, is said to be a \emph{local $C$-function\/} if, for any $x\in A$, there exists a neighborhood $B$ of $x$ in the topological space $(A, \tau_A)$, where $\tau_ A$ is the topology induced in $A$ by $\tau_C$, and a function $g\in C$ such that $ g|B=f|B$. The set of all local $C$-functions on $A$ is denoted by $C_A$ . Obviously, $C\subset C_V$. If $C=C_V$, we say that the family $C$ is \emph{closed with respect to localization}.
\begin{Lemma}
$\mathrm{Spec}\cal{A}$ and $\mathrm{Spec}C^{\infty }(\Gamma )$ are bijective.
\end{Lemma}

\noindent
{\it Proof.} The algebra $\mathcal{A}_{\Gamma }$ coincides with the algebra $C^{\infty }(\Gamma )$. Indeed, the inclusion $\mathcal{A}_{\Gamma } \subset C^{\infty }(\Gamma )$ follows from the obvious inclusion $\mathcal{A} \subset C^{\infty }(\Gamma )$. To show that $C^{\infty }(\Gamma ) \subset \mathcal{A}_{\Gamma }$, it is enogh to notice that any function $f \in C^{\infty }(\Gamma )$ is a local $\mathcal{A}$-function. Indeed, for any $\gamma \in \Gamma $ there exists a function $\varphi \in \mathcal{A}$ and an open neigborhood $U \ni \gamma $ such that $\varphi |U \equiv 1$. Then $f|U = f\varphi |U$ with $f\varphi \in \mathcal{A}$. This demonstrates that $f \in \mathcal{A}_{\Gamma }$. Therefore, $\mathcal{A}_{\Gamma } = C^{\infty }(\Gamma )$, and evidently $\mathrm{Spec}\mathcal{A}_{\Gamma } = \mathrm{Spec}C^{\infty }(\Gamma )$.

Now, we show that $\mathrm{Spec}\mathcal{A}_{\Gamma }$ is bijective with $\mathrm{Spec}\mathcal{A}$. We define the bijection $J: \mathrm{Spec}\mathcal{A} \rightarrow \mathrm{Spec}\mathcal{A}_{\Gamma }$ by
\[
J(\chi )(f) = \frac{\chi (fa)}{\chi (a)}
\]
where $f \in \mathcal{A}_{\Gamma }$ and $a$ is any element of $\mathcal{A}$ such that $\chi (a) \neq 0$. Such an element does exist since the homomorphism $\chi : \mathcal{A} \rightarrow \mathbf{C}$ is nonzero (zero homomorphisms do not belong to the spectrum). 

We should now demonstrate the correctness of the above definition.
For $a_1, a_2 \in \mathcal{A}, \, f \in \mathcal{A}_{\Gamma }$ we have
\[
\chi (a_1a_2f) = \chi (a_1 (a_2f)) = \chi(a_1) \cdot \chi (a_2f) = \chi (a_1f)\cdot \chi(a_2).
\]
Hence
\[
\frac{\chi(a_1f)}{\chi(a_1)} = \frac{\chi(a_2f)}{\chi(a_2)} 
\]
(we assume that $\chi (a_1) \neq 0$ and $\chi (a_2) \neq 0$). This shows the independence of the definition of $J$ of the choice of $a \in \mathcal{A}$. It is easy to see that $J$ is a bijection.

To end the proof we notice that the inverse mapping $J^{-1}: \mathrm{Spec}\mathcal{A}_{\Gamma } \rightarrow \mathrm{Spec}\mathcal{A}$ is given by $J^{-1}(\chi ) = \chi |\mathcal{A}$.

As a corrolary from the above Lemma we have the following sequence of bijections
\[
\mathrm{Spec}\mathcal{A} = \mathrm{Spec}C^{\infty }(\Gamma ) \simeq \Gamma.
\]
In this way, $\Gamma $ is recovered from \calA .

To recover space-time $M$ from the algebra $(\mathcal{A}, \cdot)$ we must perform the following steps:
(1) we go from \calA \ to $\mathcal{A}_{\Gamma } = C^{\infty }(\Gamma )$,
(2) we notice that $\mathcal{A}_{\Gamma }$ contains the subalgebra $(d \circ \pi_M)(C^{\infty }(M))$, and
(3) that $\mathcal{A}_{proj}$ is isomorphic with $C^{\infty }(M)$. Therefore,
\[
\mathrm{Spec}\mathcal{A}_{proj} \simeq \mathrm{Spec}C^{\infty }(M) \simeq M.
\]

To recover space-time $M$ from the algebra of random operators $(\mathcal{M}_0, \cdot )$ \ it is enough to notice that
\[
\mathrm{Spec}\mathcal{A} = \pi^*(\mathrm{Spec}\mathcal{M}_0)
\]
which is given by $ \pi^*(\chi ) = \chi \circ \pi $.

So far we were dealing with the commutative case. The only new thing was that the usual differential geometry can be done both in terms of a functional algebra on the groupoid and in terms of its representation on the algebra of random operators. Much more interesting is the noncommutative case, i.e., when we consider the algebra $(\mathcal{M}_0, \circ )$ instead of $(\mathcal{M}_0, \cdot )$. To deal with this case, let us first consider the full von Neumann algebra $\mathcal{M} = (\mathcal{M}_0)''$.

Let $\mathcal{M}_*$ be the predual of \calM , i.e., $\mathcal{M}_*$ is a unique Banach space such that  $\mathcal{M}$ is isomorphic to the Banach dual of $\mathcal{M}_*$. Let further $r \in \mathcal{M}$ and $ \varphi \in \mathcal{M}_{*}$. Then
\[
\varphi (r) = \mathrm{Tr}(\hat{\rho }(p) \circ r(p))
\]
where $\hat{\rho }$ is the density operator and $p \in E$. In this interpretation, elements of \calM \ are functions on $\mathcal{M}_*$; they are of the form $\hat{r}(\varphi ) = \varphi (r)$. In this sense, $\mathcal{M}_*$ can be regarded as the {\it virtual spectrum} of \calM . Moreover, the functionals from $\mathcal{M}_*$ are well defined on \calMM ; they are of the form $\varphi |\mathcal{M}_0$. In this sense, $\mathcal{M}_*$ can be regarded as the virtual spectrum of the algebra \calMM . We have used the term ``virtual'' to emphasize the fact that the ``true spctrum'' of \calMM \ is much bigger than $\mathcal{M}_*$ (by narrowing spectrum we do not diminish the number of functions), and $\mathcal{M}_*$ as a noncommutative algebra is not defined on any space, understood in the traditional way. However, the remarkable circumstance is that the generalized general relativity can be done in such a virtual space.

\section{Friedman World Model in Terms of Random Operators}
Im this section, to give some ``flesh'' to the above analysis, we consider the closed Friedman cosmological model as it is represented in the space of random operators. The noncommutative version of this model (in terms of the algebra \calA \ and its derivations) was presented in \cite[Sect. 5]{Conceptual}; and here we give only basic steps of the relevant computations.

Spacetime of this model is $M=\{(\eta ,\chi ,\theta ,\varphi ):\eta\in (0,T),(\chi ,\theta ,\varphi)\in
S^3\}=(0,T)\times S^3,$ where $(0,T)\subset \mathbf{R}$, and its metric is given by
\[
ds^2=R^2(\eta )(-d\eta^2+d\chi^2+\sin^2\chi (d\theta^ 2+\sin^2\theta
d\varphi^2)).
\]
The model has the initial singularity characterized by: $R^2(\eta )\rightarrow 0$ as $%
\eta\rightarrow 0$, and the final singularity characterized by: $R^2(\eta )\rightarrow 0$
as $\eta\rightarrow T$.

The total space of the frame bundle $\pi_M: E \rightarrow M$ is
\[
E=\{(\eta ,\chi ,\theta ,\varphi,\lambda ):(\eta ,\chi ,\theta ,\varphi )\in
M,\lambda\in \mathbf{R}\}=M\times \mathbf{R},
\]
and the structural group 
\[
G=\{\left(%
\begin{array}{cccc}
\cosh t & \sinh t & 0 & 0 \\
\sinh t & \cosh t & 0 & 0 \\
0 &  0 &  1 &  0 \\
0 &  0 &  0 &  1 \\
\end{array}
\right),\, t\in \mathbf{R}\}.
\]
\par
The space of the groupoid is given by
\[
\Gamma =\{(\eta ,\chi ,\theta ,\varphi ,\lambda_1,\lambda_
2):\lambda_1,\lambda_2\in \mathbf{R}\}.
\]
The $Z$-submodule $ V=V_1\oplus V_2$ of
horizontal and vertical derivations of the algebra $\mathcal{A}$ carries the metric
\begin{eqnarray*}
ds^2 &  = &  -R^2(\eta )d\eta^2+ \mbox{}R^2(\eta )d\chi^2+R^2(\eta )\sin^
2(\chi )d\theta^2+ \\
&  &  R^2(\eta )\sin^2(\chi )\sin^2(\theta )d\varphi^2+d\lambda^2.
\end{eqnarray*}

Computations of the Einstein operator give
\[\mathbf{G} = G^c_{\phantom{c}d}=%
\mathrm{diag}\{B,h,h,h,q\}
\]
where $$B = -3\frac 1{R^2(t)}-3\frac {R^{\prime}{}^2(t)}{R^4(t)},$$ 
$$h=\displaystyle-\frac 1{R^2(t)}+\frac {R^{\prime}{}^2(t)}{R^4( t)}-2\frac {R^{\prime\prime}(t)}{R^3(t)},$$ and 
$$q=\displaystyle-3\frac 1{R^2(t)}-3\frac {R^{\prime\prime} (t)}{R^3(t)}$$.

The regular reprsentation of the algebra \calA \ in the Hilbert space $\mathcal{H}_p = L^2(\Gamma^p) \simeq L^2({\mathbf{R}})$, where
\[
\Gamma^p = \{(\eta ,\chi ,\theta ,\varphi , \lambda_1 , \lambda ): \lambda_1 \in \mathbf{R}\},
\]
is
\[
(\pi_p(a)\psi )(\eta ,\chi ,\theta ,\varphi , \lambda_1 , \lambda ) = \int_{\mathbf{R}} a(\eta ,\chi ,\theta ,\varphi , \lambda_1 , \lambda_2) \psi (\eta ,\chi ,\theta ,\varphi , \lambda_2 , \lambda ) d\lambda_2,
\]
for $a \in \mathcal{A}, \psi \in \mathcal{H}_p$, and $r = (\pi_p(a))_{p \in E}$.

To write down Einstein operator in terms of random oparators we must first transfer derivations from the algebra \calA \ to the algebra \calMM . We make use of equation (\ref{derM}) to obtain the derivation $\hat{u}: \mathcal{M}_0 \rightarrow \mathcal{M}_0$
\[
(\hat{u}(\pi_p(a)))\psi(\lambda_1) = \int_{\mathbf{R}}(u(a))(\lambda_1, \lambda_2) \psi(\lambda_2)d\lambda_2.
\]
And the Einstein operator reads
\[
(\hat{G}(\hat{u}))(\pi_p(a))\psi(\lambda_1) = \int_{\mathbf{R}}(\mathbf{G}(u))(a)(\lambda_1, \lambda_2)\psi(\lambda_2)d\lambda_2.
\]

Now we can easily write the Einstein equation either in the traditional form, or as the eigenvalue equation for the Einstein operator, both in terms of random operators.

\end{document}